\documentclass[twocolumn,showpacs,showkeys]{revtex4}
\usepackage{graphicx}
\usepackage{bm}
\usepackage{color}
\usepackage{amsmath}
\usepackage{natbib}

\begin{document}

\title{Kinetic description of the oblique propagating spin-electron acoustic waves in degenerate plasmas}

\author{Pavel A. Andreev}
\email{andreevpa@physics.msu.ru}
\affiliation{Faculty of physics, Lomonosov Moscow State University, Moscow, Russian Federation.}

 \date{\today}

\begin{abstract}
Oblique propagation of the spin-electron acoustic waves in degenerate magnetized plasmas is considered in terms of quantum kinetics with the separate spin evolution,
where the spin-up electrons and the spin-down electrons are considered as two different species with different equilibrium distributions.
It is considered in the electrostatic limit.
Corresponding dispersion equation is derived.
Analytical analysis of the dispersion equation is performed in the long-wavelength limit to find an approximate dispersion equation describing the spin-electron acoustic wave.
The approximate dispersion equation is solved numerically.
Real and imaginary parts of the spin-electron acoustic wave frequency are calculated for different values of the parameters describing the system.
It is found that the increase of angle between direction of wave propagation and the external magnetic field reduces the real and imaginary parts of spin-electron acoustic wave frequency.
The increase of the spin polarization decreases the real and imaginary parts of frequency either.
The imaginary part of frequency has nonmonotonic dependence on the wave vector which shows a single maximum.
The imaginary part of frequency is small in compare with the real part for all parameters in the area of applicability of the obtained dispersion equation.
\end{abstract}

\pacs{73.22.Lp, 52.30.Ex, 52.35.Dm}
\keywords{spin-electron acoustic waves, spin plasmons, separate spin evolution, quantum plasmas}

\maketitle




\section{Introduction}

Spin effects are among the quantum effects studied in plasmas \cite{MaksimovTMP 2001,MaksimovTMP 2001 b,Marklund PRL07,Brodin PRL 08 g Kin,Andreev IJMP B 15 spin current,Mahajan PRL 11,Shukla RMP 11,Koide PRC 13}.
The separate spin evolution (SSE) model,
where the consideration of electrons with different spin projections as different species is assumed,
provides more detailed description of spin effects in plasmas in compare with the single fluid models of electron gas \cite{Andreev PRE 15 SEAW,Andreev AoP 15 SEAW,Andreev APL 16,Andreev PRE 16}.
Majority of papers on the SSE in quantum plasmas contain application of corresponding hydrodynamic model \cite{Andreev PRE 15 SEAW,Andreev PRE 16,Andreev 1510 Spin Current,Andreev_Iqbal PoP 16}.
There is a paper on the kinetic model with the SSE \cite{Andreev PoP 16 sep kin}.
These papers \cite{Andreev PoP kinetics 17 a,Andreev PoP kinetics 17 b} deal with the single fluid kinetic model of electrons, but different distributions for electrons with different spin projections are introduced at the derivation of the equilibrium distribution functions.

Spin effect is not a single quantum effect existing in plasmas. The quantum Bohm potential and exchange interaction are currently under consideration in electron-ion and dusty plasmas (see for instance \cite{Rozina PP 17}).
A difference between the group velocity and the energy velocity of
plasmonic waves on metal-insulator waveguides appearing due to the Bohm potential is demonstrated in Ref. \cite{Moradi PoP 17}.
A brief review of quantum kinetic theories of plasmas with fully degenerate electrons is presented in Ref. \cite{Brodin PPCF 17},
where a model considering the dynamics of the Fermi surface is described.

One of the main consequences of the SSE is the spin-electron acoustic wave (SEAW) \cite{Andreev PRE 15 SEAW}.
Its existence is related to the difference of the partial pressures of the spin-up and spin-down electrons.
If the SEAW propagate parallel to the external field it exists as a longitudinal wave.
Therefore, the spin-spin interaction and the equation describing the spin dynamics does not contribute in the properties of the SEAWs.
Waves of similar nature are studied in condensed matter physics \cite{Ryan PRB 91}, \cite{Agarwal PRB 14}.

A significant influence of the spin polarization on the amplitude and width of the acoustic dark soliton (presented in terms of the scalar potential of the electromagnetic field) is found for the magnetized electron-ion plasmas in Ref. \cite{Ahmad PP 16}.
It is demonstrated that the increase of the spin polarization leads to the increase of the module of the soliton amplitude.
The Raman three-wave interaction for the pump wave (O-mode), sideband Shear
Alfven wave, and the electron plasma perturbations, is considered in terms SSE-QHD equations.
It is found that the nonlinear growth rate is suppressed due to
the spin effects \cite{Shahid PP 17}.
Ion-acoustic shock wave propagation in dense magnetized plasmas with relative density
effects of spin-up and spin-down degenerate electrons is studied in Ref. \cite{Hussain PP 17},
where the transition from shock
with oscillatory trails at its wave fronts to the monotonic shock structure is studied.
The parametric role of the spin density
polarization ratio is described.

Kinetic models describe details of system evolution in the momentum space.
It reveals in a description of the Bernstein modes in the magnetized plasmas.
Kinetic models explain the Landau damping of plasma waves.
Electron Bernstein modes are waves with the frequencies close to the the harmonics of the electron cyclotron frequencies $n\mid\Omega_{e}\mid$.
They exist at the wave propagation perpendicular to the external magnetic field.
They disappear at the changing of the propagation direction towards direction parallel to the external field.
The transverse part of the electric field in the Bernstein is small in compare with longitudinal part.
Hence, the Bernstein modes can be described in the electrostatic approximation.
There are the cyclotron waves which are purely transverse waves.

Spin dynamics together with the anomalous part of the magnetic moment of electron leads to the fine structure of the cyclotron waves.
It reveals in the splitting of each cyclotron wave on three closely located branches of spectrum.
The kinetic model of degenerate electron gas presented in Refs. \cite{Andreev PoP kinetics 17 a}, \cite{Andreev PoP kinetics 17 b} does not show influence of the spin dynamics on the Bernstein mode spectrum.
Moreover, there is an extra spin caused transverse wave which is called the zeroth cyclotron wave \cite{Brodin PRL 08 g Kin}, \cite{Andreev PoP kinetics 17 b}. This wave has frequencies of order of $0.001\Omega_{e}$, where $\Omega_{e}$ is the electron cyclotron frequency. Different kinetic models of spin-1/2 quantum plasmas can be found in Refs. \cite{Hurst EPJD 14}, \cite{Zamanian EPJD 15}, \cite{Lundin PRE 10}.

These results are based on the trivial equilibrium distribution functions of the spin-1/2 particles, where $x-$ and $y-$ projections of the spin distribution function are equal to zero. The first step in the analysis of the quantum kinetic model with the non-trivial equilibrium distribution functions is presented in Ref. \cite{Andreev 1705}.

This paper is organized as follows.
In Sec. II the kinetic equations are presented.
In Sec. III the dispersion equation for oblique propagating longitudinal waves appearing at the solution of the linearized kinetic equation is presented.
Its simplified forms for waves propagating parallel or perpendicular to the external magnetic field are presented either.
In Sec. IV the spectrum of the SEAWs propagating parallel to the external magnetic field is discussed.
In Sec. V different regimes for the oblique propagation of the SEAWs are described.
In Sec. VI a brief summary of obtained results is presented.

\section{Quantum kinetic model for spin-1/2 plasmas}

Basic description of the real and imaginary parts of SEAW spectrum can be done in the electrostatic approximation which does not require equation for the evolution of the spin-distribution function. Hence, equations for the scalar distribution functions are presented below.

Kinetic equations for the scalar distributions of electrons with s-spin projection are derived in Ref. \cite{Andreev PoP 16 sep kin}
$$\partial_{t}f_{e,s}+\textbf{v}\cdot\nabla_{\textbf{r}}f_{e,s} +q_{e}\biggl(\textbf{E}+\frac{1}{c}[\textbf{v},\textbf{B}]\biggr)\cdot\nabla_{\textbf{p}}f_{e,s}$$
$$\pm\gamma_{e}\nabla B^{z} \cdot\nabla_{\textbf{p}} f_{e,s}
+\frac{\gamma_{e}}{2}(\nabla B_{x} \cdot\nabla_{\textbf{p}} S_{e,x}$$
\begin{equation}\label{SUSDKin kin eq f e up}  +\nabla B_{y} \cdot\nabla_{\textbf{p}} S_{e,y})=\pm\frac{\gamma_{e}}{\hbar}[S_{e,x} B_{y}-S_{e,y} B_{x}] .\end{equation}
These equations contain $x-$ and $y-$projections of the vector (spin) distribution functions of all electrons ($S_{e,x}$ and $S_{e,y}$).

Kinetic equations are coupled with the electrostatic limit of Maxwell equations
\begin{equation}\label{SUSDKinII div E} \nabla\cdot \textbf{E}=4\pi e(n_{i}-n_{eu}-n_{ed}),\end{equation}
and
$\nabla\times \textbf{E}=0$, where
$n_{es}=\int f_{es}(\textbf{r},\textbf{p},t)d\textbf{p}$.

The step functions of different width are applied for the equilibrium distribution functions of degenerate electrons with different spin projections.
\begin{equation}\label{SUSDKinII equilib distrib el pol} f_{0s}=\frac{1}{(2\pi\hbar)^{3}}\Theta(p_{Fs}-p),\end{equation}
where $p_{Fs}=(6\pi^{2}n_{0s})^{\frac{1}{3}}\hbar$, and $s=u$, or $d$.

The spin polarization is caused by the external magnetic field. It leads to the traditional equation for the spin polarization $\eta=|n_{u}-n_{d}|/(n_{u}+n_{d})=\tanh(\gamma B_{ext}/\varepsilon_{Fe})$, where $\varepsilon_{Fe}=(3\pi^{2}n_{0e})^{2/3}\hbar^{2}/2m$ is the Fermi energy.
The spin polarization can be caused by the inner interaction in the material as it happens inside of domains in ferromagnetic materials.
In this case, an effective magnetic field $B_{eff}$ combines with the external magnetic field $B_{ext}$.

Presented here kinetic equations are obtained by the method of quantum kinetics \cite{Andreev PoP 16 sep kin} which is a generalization of the many-particle quantum hydrodynamic theory (MPQHDT) \cite{MaksimovTMP 2001}, \cite{MaksimovTMP 2001 b}.
The MPQHDT aims a representation of exact microscopic quantum dynamics of the physical system in terms of collective variables, such as the concentration $n(\textbf{r},t)$ (number of particles in the vicinity of the point of space), momentum density (current) $\textbf{j}(\textbf{r},t)$, energy, pressure tensor, quantum stress tensor for the particles with the short range interaction \cite{Andreev PRA08}, etc.
The MPQHDT gives evolution of exact functions. This evolution happens in accordance with the chosen Hamiltonian modeling considering system.
However, the MPQHDT makes no averaging.
Replacement of the many-particle $3N-$dimensional wave function with the collective variables requires almost infinite number of the collective variables.
It gives obviously impossible task.
Therefore, this is the step, where some approximations are required.
Part of information about system is lost if a finite number of collective variables is chosen.
However, it leads to a solvable model giving required properties of the system.
Quantum mechanics is interpreted in terms of probability. However, presented description in terms of collective variables provides no additional probabilities which can be met in many works on statistical physics, where canonical ensembles are introduced via "classical" probability (see for instance recent work on quantum plasma \cite{Moldabekov arxiv 17}).
From technical point of view the introduction of the additional "classical" probability can be useful.
However, it clouds our understanding of relation between two classes of models: wave function (microscopic model), where focus is on the dynamics of quantum particles, and collective functions (mesoscopic model), where focus is on the local evolution of characteristics of all system (or its mesoscopic parts).

\begin{figure}
\includegraphics[width=8cm,angle=0]{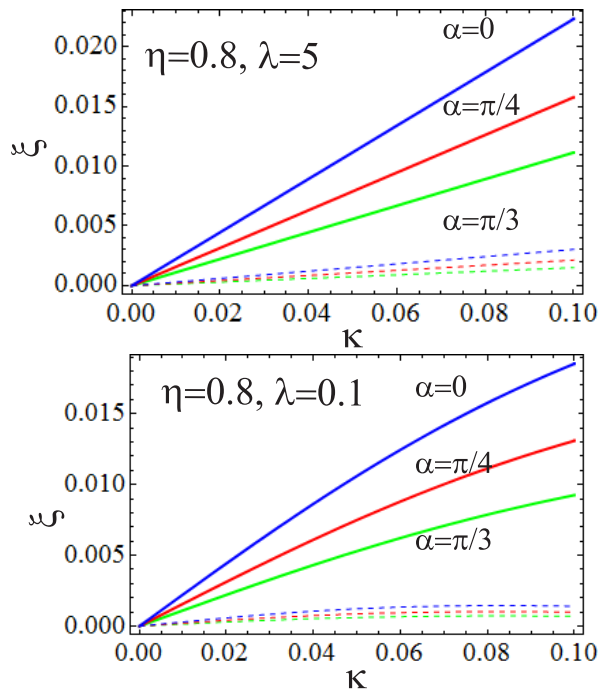}
\caption{\label{SUSDKinII 01} (Color online)
The figure shows the real ($Re\xi$, continuous lines) and minus imaginary ($-Im\xi$, dashed lines) parts
of the frequency of the SEAWs for three directions of wave propagation.
Angles $\alpha=0$, $\alpha=\pi/4$, $\alpha=\pi/3$ are shown in figure near corresponding curves.
Chosen value of spin polarization $\eta$ and ration between
the Langmuir frequency $\omega_{Le}$ and the cyclotron frequency $\Omega_{e}$
(parameter $\lambda$) are shown in the figure either.
All figures show that the increase of the angle $\alpha$ monotonically decreases real and imaginary parts of the frequency.
Upper (blue) lines correspond to $\alpha=0$. Middle (red) lines correspond to $\alpha=\pi/4$.
Lower (green) lines correspond to $\alpha=\pi/3$.}
\end{figure}

\begin{figure}
\includegraphics[width=8cm,angle=0]{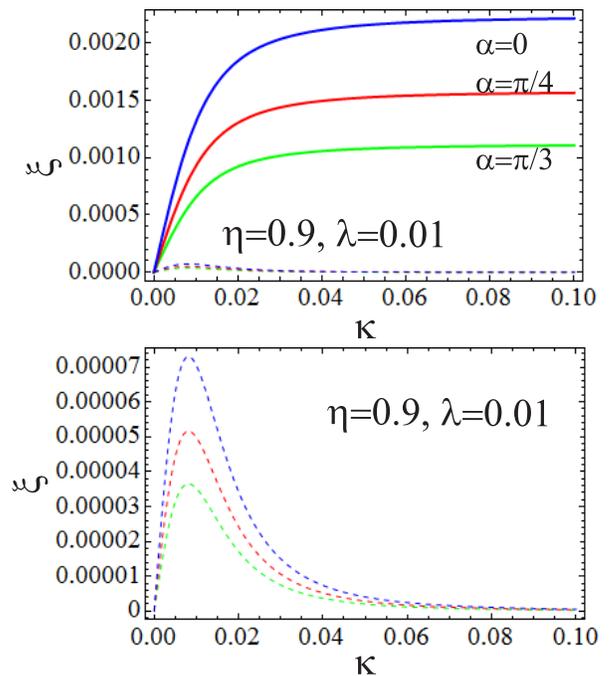}
\caption{\label{SUSDKinII 02} (Color online) The figure showsthe real ($Re\xi$, continuous lines) and minus imaginary ($-Im\xi$, dashed lines) parts
of the frequency of the SEAWs for three directions of wave propagation for $\eta=0.9$ and $\lambda=0.01$.
Here and below, the details of the imaginary parts of frequency are shown in additional figure.} \end{figure}

\begin{figure}
\includegraphics[width=8cm,angle=0]{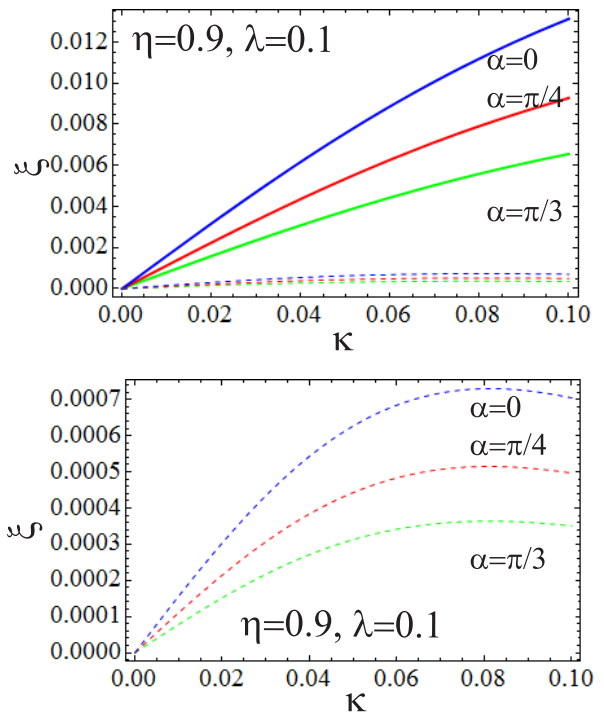}
\caption{\label{SUSDKinII 03} (Color online) The figure shows the real ($Re\xi$, continuous lines) and minus imaginary ($-Im\xi$, dashed lines) parts
of the frequency of the SEAWs for three directions of wave propagation for $\eta=0.9$ and $\lambda=0.1$.} \end{figure}

\section{Linearized kinetic equations and solutions for distribution functions}

Equilibrium condition can be described by the non-zero scalar distribution functions $f_{0eu}$, $f_{0ed}$, and an external magnetic field $\textbf{B}_{ext}=B_{0}\textbf{e}_{z}$, but $\textbf{E}_{0}=0$.
Assuming that perturbations are monochromatic which means that functions $\delta f_{eu}$, $\delta f_{ed}$, $\delta \textbf{E}$, can be presented as amplitude $F_{Au}$, $F_{Ad}$, $\textbf{E}_{A}$, correspondingly, multiplied by $e^{-\imath\omega t+\imath \textbf{k} \textbf{r}}$,
we obtain a set of linear algebraic equations relatively to $F_{Au}$, $F_{Ad}$, $\textbf{E}_{A}$.
Condition of existence of nonzero solutions for amplitudes of perturbations gives us a dispersion equation.

As it is demonstrated in Ref. \cite{Andreev PoP 16 sep kin} (see equation 41) by the standard methods of kinetic equation solution, the dispersion equation for the longitudinal small amplitude perturbations oblique propagating in magnetized degenerate plasmas with the SSE has the following form:
$$1+\frac{8\pi^{2}e^{2}}{k}\Biggl\{\sum_{s=u,d}\sum_{n=-\infty}^{+\infty}\frac{p_{Fs}^{2}}{(2\pi\hbar)^{3}}\int_{0}^{\pi}\sin\theta d\theta\times$$
\begin{equation}\label{SUSDKinII permittivity} \times \frac{J_{n}^{2} (\frac{k_{x}v_{Fs}}{\Omega_{e}}\sin\theta )}{k_{z}v_{Fs}\cos\theta-\omega+n\Omega_{e}}
\Biggl[\cos\alpha\cos\theta +\frac{n \Omega_{e}}{kv_{Fs}}\Biggr]\Biggr\}=0, \end{equation}
where $v_{z}=v\cos\theta$, $v_{\perp}=v\sin\theta$, $k_{z}=k\cos\alpha$, $k_{x}=k\sin\alpha$, and $J_{n}(x)$ are the Bessel functions.
The derivation of the dispersion equation (\ref{SUSDKinII permittivity}) applied the standard techniques \cite{Landau v10}, \cite{Rukhadze book 84}.

Dispersion equation (\ref{SUSDKinII permittivity}) simplifies to
\begin{equation}\label{SUSDKinII DE General for Parallel Pr} 1+\frac{8\pi^{2}e^{2}}{k^{2}}\sum_{s=u,d}\frac{m_{e}^{2}v_{Fs}}{(2\pi\hbar)^{3}}\biggl(2+\omega\int_{0}^{\pi}\frac{\sin\theta d\theta}{kv_{Fs}\cos\theta-\omega}\biggr)=0\end{equation}
at the wave propagation parallel to the external magnetic field
$\textbf{k}\parallel \textbf{B}_{0}$, $\alpha=0$, $k_{x}=0$, $k_{z}=k$, $J_{n}(0)=0$ if $n\neq0$, and $J_{0}(0)=1$.
This regime is considered in Ref. \cite{Andreev PoP 16 sep kin}, where the following equation is derived and studied:
\begin{equation}\label{SUSDKinII DE Longit waves DI}  1=\sum_{a=u,d}\frac{3}{2}\frac{\omega_{La}^{2}}{v_{Fa}^{2}k^{2}}\biggl(\frac{\omega}{kv_{Fa}}\ln\frac{\omega+kv_{Fa}}{\omega-kv_{Fa}}-2\biggr). \end{equation}
Equation (\ref{SUSDKinII DE Longit waves DI}) describes two wave solution: the Langmuir wave and the SEAW.

Consider opposite limit, the regime of wave propagation perpendicular to the external magnetic field.
Thus, dispersion equation (\ref{SUSDKinII permittivity}) simplifies to
\begin{equation}\label{SUSDKinII permittivity for perpendicular} 1+\frac{8\pi^{2}e^{2}}{k^{2}}
\sum_{s=u,d}\sum_{n=-\infty}^{+\infty} \frac{m^2 v_{Fs}}{(2\pi\hbar)^{3}} \frac{I_{n,s}n\Omega_{e}}{n\Omega_{e}-\omega}
=0, \end{equation}
where $I_{n,s}(k_{x}v_{Fs}/\Omega_{e})=\int_{0}^{\pi} J_{n}^{2}(\frac{k_{x}v_{Fs}}{\Omega_{e}}\sin\theta)\sin\theta d\theta$ parameters which do not depend on the frequency $\omega$.
Equation (\ref{SUSDKinII permittivity for perpendicular}) shows that the SSE does not change number of solution of the dispersion equation in the regime of perpendicular propagation.
Hence, the number of waves is not affected by the SSE, but form of spectrum is modified.

\section{Propagation of SEAW parallel to the external field}

Consider results for the SEAW propagation parallel to the external field. Following Ref. \cite{Andreev PoP 16 sep kin} consider the regime of intermediate phase velocities $kv_{Fu}\ll\omega\ll kv_{Fd}$.
Hence, equation (\ref{SUSDKinII DE Longit waves DI}) can be simplifies at the application of equation (\ref{SUSDKinII log small omega}) for the term describing spin-down electrons
and equation (\ref{SUSDKinII log large omega}) with $n=0$ for the spin-up electrons.
Keeping major term in the expansion find the following simplified dispersion equation
$$1+3 \frac{\omega_{Ld}^{2}}{k^{2}v_{Fd}^{2}}\biggl(1+\frac{\pi}{2}\imath\frac{\omega}{kv_{Fd}}-\frac{\omega^{2}}{k^{2}v_{Fd}^{2}}\biggr)$$
\begin{equation}\label{SUSDKinII SEAW DI DE} = \frac{\omega_{Lu}^{2}}{\omega^{2}}\biggl(1+\frac{3}{5}\frac{k^{2}v_{Fu}^{2}}{\omega^{2}}\biggr). \end{equation}

In considering regime, real part of the SEAWs spectrum is obtained from equation (\ref{SUSDKinII SEAW DI DE})
\begin{equation}\label{SUSDKin SEAW DD I order} \omega_{R}^{2}=\frac{\omega^{2}_{Lu}}{1+3\frac{\omega_{Ld}^{2}}{k^{2}v_{Fd}^{2}}}. \end{equation}
In the long-wavelength limit it simplifies to the linear spectrum $\omega_{R}=(\omega_{Lu}v_{Fd}/\sqrt{3}\omega_{Ld})k$.

Equation (\ref{SUSDKinII SEAW DI DE}) has an imaginary part.
Hence, the frequency is complex that leads to the Landau damping of the SEAW:
\begin{equation}\label{SUSDKinII SEAW DI LD} \omega_{Im}=\frac{1}{2}\omega_{R} \frac{\frac{3\pi}{2}\frac{\omega_{Ld}^{2}}{k^{2}v_{Fd}^{2}}\frac{\omega_{R}}{kv_{Fd}}}{1+3\frac{\omega_{Ld}^{2}}{k^{2}v_{Fd}^{2}} -3\frac{\omega_{Ld}^{2}\omega_{R}^{2}}{k^{4}v_{Fd}^{4}}}. \end{equation}

At $\omega_{Ld}^{2}\gg k^{2}v_{Fd}^{2}$, we find a simplification of equation (\ref{SUSDKinII SEAW DI LD}) to the following form $\omega_{Im}=\frac{\pi}{4}\frac{\omega_{R0}}{kv_{Fd}}\omega_{R0}\ll\omega_{R0}$. It demonstrates that the damping is small. Hence, it is meaningful to deal with propagation of the SEAWs.

\section{Oblique propagation of the SEAWs in the long-wavelength limit}

Apply equation (\ref{SUSDKinII permittivity}) for the analysis of the oblique
propagating waves
and focus on the long-wavelength limit.
To this end, the Bessel functions are expanded in to the Taylor series.
Consider the first term of the series only: $J_{n}(x)\approx(x/2)^{n}/n!$.

\begin{figure}
\includegraphics[width=8cm,angle=0]{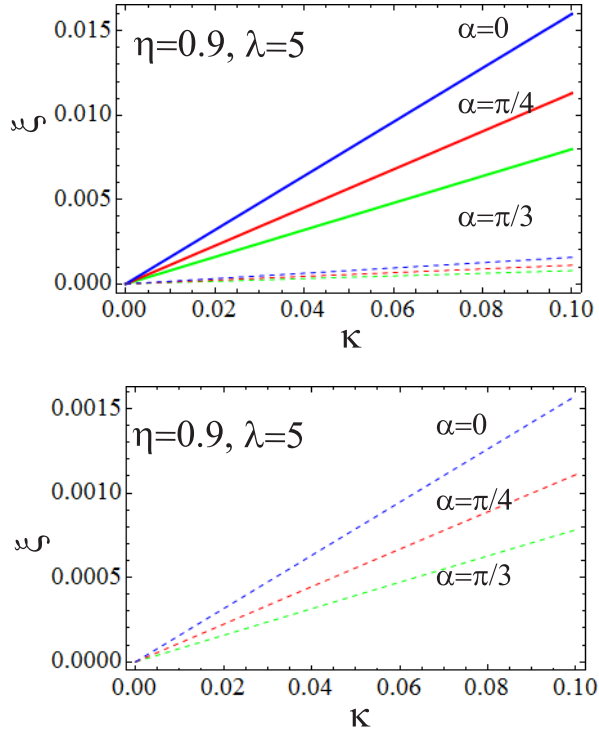}
\caption{\label{SUSDKinII 04} (Color online) The figure shows the real ($Re\xi$, continuous lines) and minus imaginary ($-Im\xi$, dashed lines) parts
of the frequency of the SEAWs for three directions of wave propagation for $\eta=0.9$ and $\lambda=5$.} \end{figure}

\begin{figure}
\includegraphics[width=8cm,angle=0]{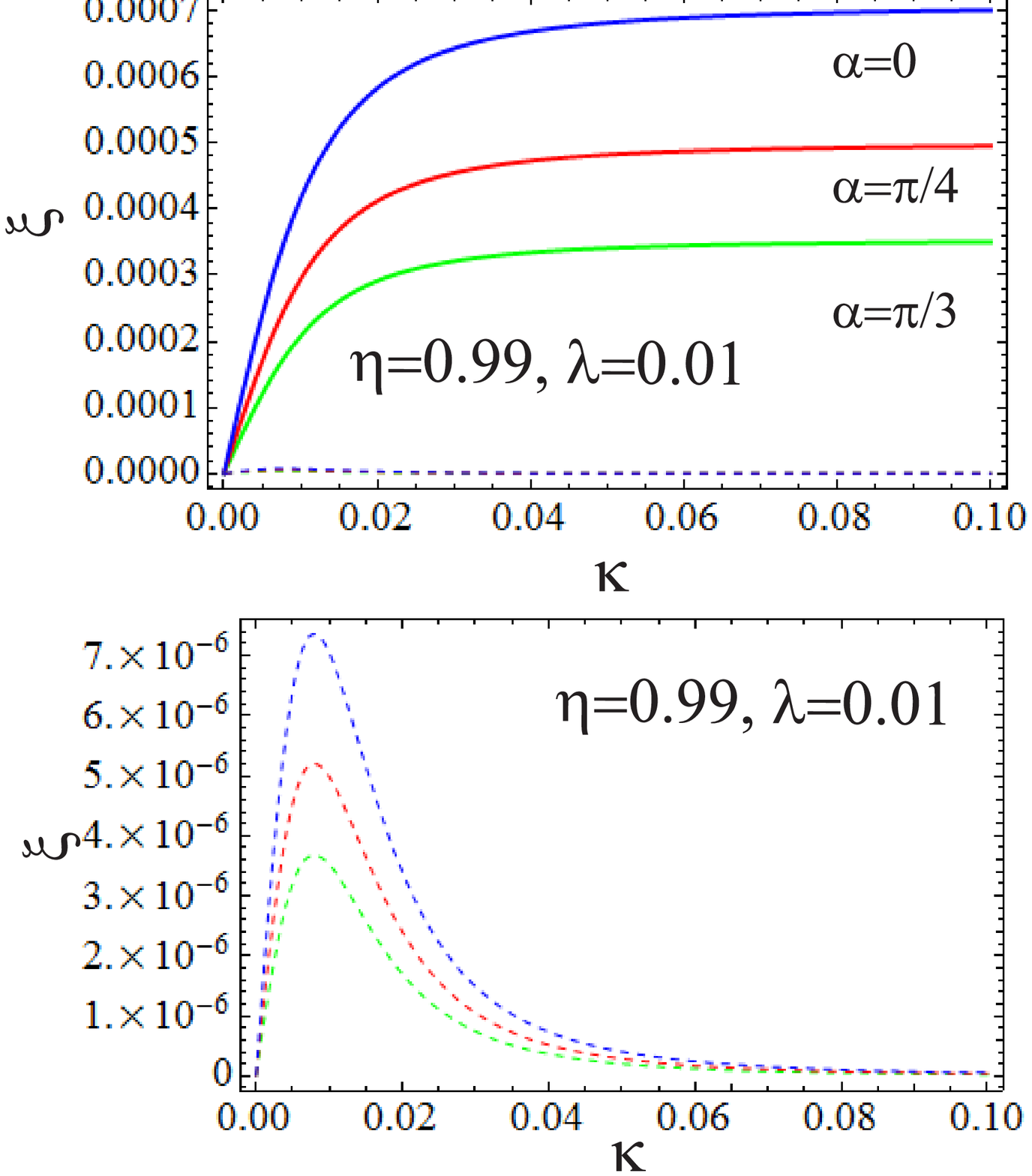}
\caption{\label{SUSDKinII 05} (Color online) The figure shows the real ($Re\xi$, continuous lines) and minus imaginary ($-Im\xi$, dashed lines) parts
of the frequency of the SEAWs for three directions of wave propagation for $\eta=0.99$ and $\lambda=0.01$.} \end{figure}

\begin{figure}
\includegraphics[width=8cm,angle=0]{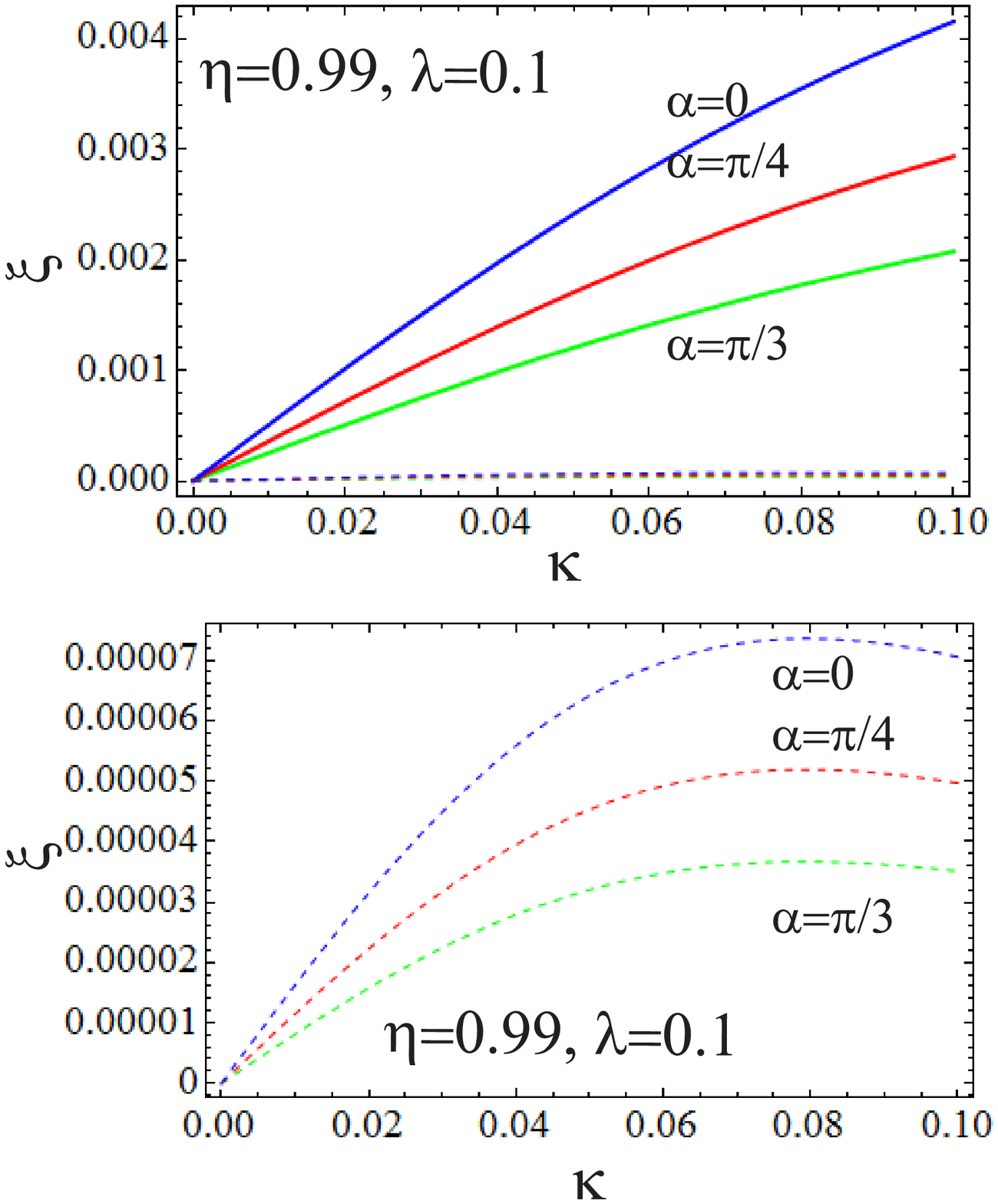}
\caption{\label{SUSDKinII 06} (Color online) The figure shows the real ($Re\xi$, continuous lines) and minus imaginary ($-Im\xi$, dashed lines) parts
of the frequency of the SEAWs for three directions of wave propagation for $\eta=0.99$ and $\lambda=0.1$.} \end{figure}

\begin{figure}
\includegraphics[width=8cm,angle=0]{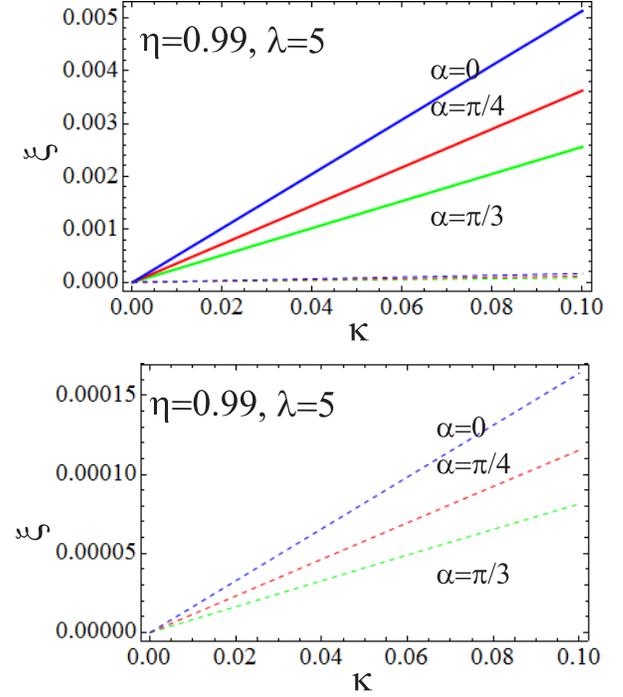}
\caption{\label{SUSDKinII 07} (Color online) The figure shows the real ($Re\xi$, continuous lines) and minus imaginary ($-Im\xi$, dashed lines) parts
of the frequency of the SEAWs for three directions of wave propagation for $\eta=0.99$ and $\lambda=5$.} \end{figure}

After expansion of the Bessel functions and integration over the angle $\theta$, equation (\ref{SUSDKinII permittivity}) reappears in a huge form.
Corresponding results are presented in Appendix A, where the terms with
numbers $n\in[-4,4]$ are included.

The regime of further expansion of logarithmic functions presented by equation (\ref{SUSDKinII App A main Eq}) in Appendix A depends on the value of frequency.

Next, let us consider a regime, where the frequency is close to a harmonic of the cyclotron frequency: $a_{u}\ll \mid n_{0}-\xi\mid\ll a_{d}$ for a fixed $n_{0}$, hence $a_{u}\leq a_{d}\ll \mid n-\xi\mid$ for all $n\neq n_{0}$. It requires considerable difference between $a_{u}$ and $a_{d}$ which can be reached at the large spin polarizations $\eta\geq0.9$.

If $n_{0}=0$, the following dispersion equation is found as the corresponding limit of equation (\ref{SUSDKinII App A main Eq})
$$1-\frac{1}{2}(1-\eta)\lambda^{2}\frac{\cos^{2}\alpha}{\xi^{2}}$$
$$+\frac{3}{4}(1+\eta)\frac{\lambda^{2}}{a_{d}^{2}}\biggl(2-\frac{2\xi^{2}}{a_{d}^{2}\cos^{2}\alpha}+\pi\imath\frac{\xi}{a_{d}\cos\alpha}  \biggr)$$
\begin{equation}\label{SUSDKinII OblPr no Ln n=0}
+\sum_{s}3\lambda^{2}\frac{n_{0s}}{n_{0e}}\sum_{n=1}^{5}\frac{a_{s}^{2n-2}\sin^{2n}\alpha}{c_{n}[n!]^{2}}=0,\end{equation}
where $c_{1}=3$, $c_{2}=3\cdot5$, $c_{3}=2\cdot5\cdot7$, $c_{4}=5\cdot7\cdot9$, $c_{5}=2\cdot7\cdot9\cdot11$, $\xi=\omega/\mid\Omega_{e}\mid$, $\lambda=\omega_{Le}/\mid\Omega_{e}\mid$, $\kappa=kv_{Fe}/\mid\Omega_{e}\mid$, $a_{u}=(1-\eta)^{1/3}\kappa$, $a_{d}=(1+\eta)^{1/3}\kappa$, and $\xi\rightarrow 0$ is placed in the last term.
Presented here consideration of $n=0$ leads to the oblique propagating SEAW described above for the regime of propagation parallel to the external magnetic field.

Derivation of the dispersion equation (\ref{SUSDKinII OblPr no Ln n=0}) is performed up to the $n$-th harmonic.
However, there is no need in extra calculation for generalization of equation (\ref{SUSDKinII OblPr no Ln n=0}) for estimation of the higher harmonic contribution in the spectrum of the SEAWs.

Let us compare different terms in the last group of terms in equation (\ref{SUSDKinII OblPr no Ln n=0}).
It can be seen that the increase of the number increases the denominator $\sim c_{n}[n!]^{2}$, where $c_{n}$ increases with the increase of $n$ as it can be seen from the explicit form of coefficients $c_{n}$. Moreover, each term is proportional to $a_{s}^{2n-2}\sin^{2n}\alpha$. Obviously, multiplier $\sin^{2n}\alpha\leq1$ is small and multiplier $a_{s}^{2n-2}$ is small either, since the long-wavelength is considered. Therefore, terms with larger numbers $n$ give negligible contribution in the SEAW spectrum.

Dropping the second and third terms in the third group of terms in equation (\ref{SUSDKinII OblPr no Ln n=0}) and keeping the first term in the last group of terms in equation (\ref{SUSDKinII OblPr no Ln n=0}), we find an analytical solution for the real part of frequency of the oblique propagating SEAWs:
\begin{equation}\label{SUSDKinII OblPr no Ln n=0 Re sol} \xi^{2}=\frac{1}{2}\frac{(1-\eta)\lambda^{2}\cos^{2}\alpha}{1+\lambda^{2}\sin^{2}\alpha+\frac{3}{2}(1+\eta)\frac{\lambda^{2}}{a_{d}^{2}}}. \end{equation}
The long-wavelength limit is under consideration, so it leads to the following approximation of equation (\ref{SUSDKinII OblPr no Ln n=0 Re sol})
\begin{equation}\label{SUSDKinII OblPr no Ln n=0 Re sol} \xi^{2}=\frac{1}{3}\frac{1-\eta}{1+\eta}a_{d}^{2}\cos^{2}\alpha. \end{equation}
Including imaginary term in equation (\ref{SUSDKinII OblPr no Ln n=0}) by the iteration method, we find the following solution
\begin{equation}\label{SUSDKinII OblPr no Ln n=0 C sol} \xi^{2}=\frac{1}{3}\frac{1-\eta}{1+\eta}a_{d}^{2}\cos^{2}\alpha \biggl(1-\frac{\pi\imath}{2\sqrt{3}}\sqrt{\frac{1-\eta}{1+\eta}}\biggr). \end{equation}
These equations show approximate behavior of the SEAW frequency.

This estimation of the imaginary part is found by the iteration method assuming that the imaginary term in equation (\ref{SUSDKinII OblPr no Ln n=0}) is small.
Solution (\ref{SUSDKinII OblPr no Ln n=0 C sol}) is in agreement with applied approximation if system has high spin polarization $1-\eta\ll1$.

Simultaneous decrease of the real and imaginary parts of frequency at the increase of angle $\alpha$ can be seen from equation (\ref{SUSDKinII OblPr no Ln n=0 C sol}).

All figures demonstrate real and imaginary parts of frequency of the SEAW.
Each of them is obtained for three directions of wave propagation
($\alpha=0$, $\alpha=\pi/4$, $\alpha=\pi/3$).
The small values $\kappa<0.1$ of the dimensionless wave vector are presented in all figures since the long-wavelength limit is considered.

Fig. (\ref{SUSDKinII 01}), with $\eta=0.8$, and different values of $\lambda$, shows that
the increase of the wave vector decreases the real and imaginary parts of the group velocity
$d\omega/d k=d\xi/d\kappa$.
It is noticeable at the small $\lambda$.
Moreover, the increase of angle $\alpha$  leads to simultaneous decrease of
real and imaginary parts of frequency in accordance with approximate solution (\ref{SUSDKinII OblPr no Ln n=0 C sol}).

Analysis of Figs. (\ref{SUSDKinII 01})-(\ref{SUSDKinII 07}) shows that the decrease of parameter $\lambda$ at the fixed spin polarization $\eta$ leads to the decrease of the real $Re\xi$ and minus imaginary $-Im\xi$ parts of frequency.
Small change of frequency is found at transition from $\lambda=5$ to $\lambda=0.1$.
Relatively larger change of frequency is observed at the further transition to $\lambda=0.01$.
It is noticeable for small values of $\lambda$ ($\lambda\ll1$).
This behavior corresponds to equation (\ref{SUSDKinII OblPr no Ln n=0 Re sol}).
Decrease of the spin polarization $\eta$ at the fixed parameter $\lambda$ leads to the increase of the real $Re\xi$ and minus imaginary $-Im\xi$ parts of frequency.
It is also in accordance with equation (\ref{SUSDKinII OblPr no Ln n=0 C sol}).

Mentioned above effect that the increase of angle $\alpha$ leads to simultaneous decrease of
real and imaginary parts of frequency is correct for different values of $\eta$ and $\lambda$ as it is follows from Figs. (\ref{SUSDKinII 01})-(\ref{SUSDKinII 07}).

Fig. (\ref{SUSDKinII 01}) is obtained for the relatively small
spin polarization $\eta=0.8$, relatively chosen area of research,
where $\eta\rightarrow1$.
In this regime, the damping $\sim-Im\xi$ is rather high since $-Im\xi/Re\xi\sim0.1$.
Shifting to the area of larger spin polarization, corresponding to the area of
better applicability of equation (\ref{SUSDKinII OblPr no Ln n=0}),
we find that the damping decreases.
At $\eta=0.9$ and $\lambda=0.01$, presented in Fig. (\ref{SUSDKinII 02}),
there is maximum of $-Im\xi$
which has same location on the wave vector scale at different directions of wave propagation.
In this case, it is reached at $\kappa\approx0.01$.
For $\alpha=\pi/4$, Fig. (\ref{SUSDKinII 02}) shows
$-Im\xi(\kappa=0.01)=5\times10^{-5}$ and $Re\xi=10^{-3}$.
It gives $-Im\xi/Re\xi=2\times10^{-3}$ and shows small damping of the SEAW.
Fig. (\ref{SUSDKinII 02}) shows that the damping decrement $Im\xi$
is not monotonic function of the wave vectors.
After reaching a maximum at an intermediate value of wave vector,
the damping decrement decreases monotonically with the growth of $\kappa$.
Change of the direction of wave propagation does not make
noticeable changes in the position of the maximum of the damping decrement.

\section{Conclusion}

The real and imaginary parts of frequency of the SEAWs propagating with
arbitrary angle relatively to the external magnetic field have been derived.
The frequency has been found in the linear approximation of the quantum kinetic model with the separate spin evolution.
Regimes of small damping and large damping have been found.

The long-wavelength limit of the spectrum of SEAWs is considered numerically and analytically.
Almost linear spectrum of SEAW at small wave vectors is found.
With the increase of the wave vector the spectrum can reach a plateau.
It has been mentioned in the literature that there is a regime, where the SEAWs have small damping at the parallel propagation.
Here, it has been shown numerically for variety of parameters that the damping is small for the long-wavelength oblique propagating SEAWs.
Details of the imaginary part of spectrum are studied.
A nonmonotonic behavior of the imaginary part of frequency has been found, where single maximum has been observed.
Position of the maximum on the wave vector axis does not depend on the direction of wave propagation.
It does not show dependence on the spin polarization either, but heavily depends on the ration of the Langmuir frequency to the cyclotron frequency $\lambda$.
For instance, transition from $\lambda=0.01$ to $\lambda=0.1$ shifts the position of the maximum from $\kappa=0.01$ to $\kappa=0.8$.
However, the amplitude of the maximum depends on the spin polarization and on the direction of wave propagation.

\section{Appendix A: Intermediate form of the dispersion equation for the oblique propagating Bernstein modes in the long-wavelength regime}

Considering the long-wavelength limit of equation (\ref{SUSDKinII permittivity})
and including terms up to number $\mid n\mid=4$,
find the following form of the dispersion equation for the oblique
propagating longitudinal waves:
\begin{widetext}
$$0=1+\sum_{s}\frac{3}{2}\frac{n_{0s}}{n_{0e}}\frac{\lambda^{2}}{a_{s}^{2}}\Biggl\{ \biggl[2-\frac{\xi Ln(0)}{a_{s}\cos\alpha}\biggr]$$
$$+\biggl[\frac{a_{s}^{2}\sin^{2}\alpha}{2^{2}}         \biggl[\frac{2^{2}}{3} -\frac{\xi Ln(1)}{a_{s}\cos\alpha}\cdot \biggl(1-\frac{(1-\xi)^{2}}{a_{s}^{2}\cos^{2}\alpha}\biggr) +2\xi\frac{1-\xi}{a_{s}^{2}\cos^{2}\alpha}\biggr]\biggr]$$
$$+\biggl[\frac{a_{s}^{2}\sin^{2}\alpha}{2^{2}}         \biggl[\frac{2^{2}}{3} -\frac{\xi Ln(-1)}{a_{s}\cos\alpha}\cdot \biggl(1-\frac{(1+\xi)^{2}}{a_{s}^{2}\cos^{2}\alpha}\biggr) -2\xi\frac{1+\xi}{a_{s}^{2}\cos^{2}\alpha}\biggr]\biggr]$$
$$+\biggl[\frac{a_{s}^{4}\sin^{4}\alpha}{2^{4}(2!)^{2}} \biggl[\frac{2^{4}}{3\cdot5} -\frac{\xi Ln(2)}{a_{s}\cos\alpha}\cdot \biggl(1-\frac{(2-\xi)^{2}}{a_{s}^{2}\cos^{2}\alpha}\biggr)^{2} +2\xi\frac{2-\xi}{a_{s}^{2}\cos^{2}\alpha}\biggl(\frac{5}{3}-\frac{(2-\xi)^{2}}{a_{s}^{2}\cos^{2}\alpha}\biggr)\biggr]\biggr]$$
$$+\biggl[\frac{a_{s}^{4}\sin^{4}\alpha}{2^{4}(2!)^{2}} \biggl[\frac{2^{4}}{3\cdot5} -\frac{\xi Ln(-2)}{a_{s}\cos\alpha}\cdot \biggl(1-\frac{(2+\xi)^{2}}{a_{s}^{2}\cos^{2}\alpha}\biggr)^{2} -2\xi\frac{2+\xi}{a_{s}^{2}\cos^{2}\alpha}\biggl(\frac{5}{3}-\frac{(2+\xi)^{2}}{a_{s}^{2}\cos^{2}\alpha}\biggr)\biggr]\biggr]$$
$$+\biggl[\frac{a_{s}^{6}\sin^{6}\alpha}{2^{6}(3!)^{2}} \biggl[\frac{2^{5}}{5\cdot7} -\frac{\xi Ln(3)}{a_{s}\cos\alpha}\cdot \biggl(1-\frac{(3-\xi)^{2}}{a_{s}^{2}\cos^{2}\alpha}\biggr)^{3}  +2\xi\frac{3-\xi}{a_{s}^{2}\cos^{2}\alpha}\biggl(\frac{11}{5}-\frac{8}{3}\frac{(3-\xi)^{2}}{a_{s}^{2}\cos^{2}\alpha} +\frac{(3-\xi)^{4}}{a_{s}^{4}\cos^{4}\alpha}\biggr)\biggr]\biggr]$$
$$+\biggl[\frac{a_{s}^{6}\sin^{6}\alpha}{2^{6}(3!)^{2}} \biggl[\frac{2^{5}}{5\cdot7} -\frac{\xi Ln(-3)}{a_{s}\cos\alpha}\cdot \biggl(1-\frac{(3+\xi)^{2}}{a_{s}^{2}\cos^{2}\alpha}\biggr)^{3}  -2\xi\frac{3+\xi}{a_{s}^{2}\cos^{2}\alpha}\biggl(\frac{11}{5}-\frac{8}{3}\frac{(3+\xi)^{2}}{a_{s}^{2}\cos^{2}\alpha} +\frac{(3+\xi)^{4}}{a_{s}^{4}\cos^{4}\alpha}\biggr)\biggr]\biggr]$$
$$+\biggl[\frac{a_{s}^{8}\sin^{8}\alpha}{2^{8}(4!)^{2}} \biggl[\frac{2^{8}}{5\cdot7\cdot9} -\frac{\xi Ln(4)}{a_{s}\cos\alpha}\cdot \biggl(1-\frac{(4-\xi)^{2}}{a_{s}^{2}\cos^{2}\alpha}\biggr)^{4}  +2\xi\frac{4-\xi}{a_{s}^{2}\cos^{2}\alpha}\biggl(\frac{93}{5\cdot7}-\frac{73}{3\cdot5}\frac{(4-\xi)^{2}}{a_{s}^{2}\cos^{2}\alpha} +\frac{11}{3}\frac{(4-\xi)^{4}}{a_{s}^{4}\cos^{4}\alpha}-\frac{(4-\xi)^{6}}{a_{s}^{6}\cos^{6}\alpha}\biggr)\biggr]\biggr]$$
\begin{equation}\label{SUSDKinII App A main Eq}
+\biggl[\frac{a_{s}^{8}\sin^{8}\alpha}{2^{8}(4!)^{2}} \biggl[\frac{2^{8}}{5\cdot7\cdot9} -\frac{\xi Ln(-4)}{a_{s}\cos\alpha}\cdot \biggl(1-\frac{(4+\xi)^{2}}{a_{s}^{2}\cos^{2}\alpha}\biggr)^{4}  -2\xi\frac{4+\xi}{a_{s}^{2}\cos^{2}\alpha}\biggl(\frac{93}{5\cdot7}-\frac{73}{3\cdot5}\frac{(4+\xi)^{2}}{a_{s}^{2}\cos^{2}\alpha} +\frac{11}{3}\frac{(4+\xi)^{4}}{a_{s}^{4}\cos^{4}\alpha}-\frac{(4+\xi)^{6}}{a_{s}^{6}\cos^{6}\alpha}\biggr)\biggr]\biggr] \Biggr\},
\end{equation}\end{widetext}
where
\begin{equation}\label{SUSDKinII} Ln(n) \equiv \ln\biggl(\frac{n-\xi-a_{s}\cos\alpha}{n-\xi+a_{s}\cos\alpha}\biggr).\end{equation}

If $a_{s}\cos\alpha\ll\mid n-\xi\mid$, we find an approximate form of function $Ln(n)$:
\begin{equation}\label{SUSDKinII log large omega} Ln(n)\approx -2\sum_{l=0}^{\infty}\frac{1}{2l+1}\biggl(\frac{a_{s}\cos\alpha}{n-\xi}\biggr)^{2l+1}\end{equation}
It is enough including $l$ up to $l=n$ in each group of terms.

At $\mid n-\xi\mid\ll a_{s}\cos\alpha$, we obtain another well-known expansion
\begin{equation}\label{SUSDKinII log small omega} Ln(n)\approx -\pi\imath-2\frac{n-\xi}{a_{s}\cos\alpha}\biggl(1+\frac{1}{3}\frac{(n-\xi)^{2}}{a_{s}^{2}\cos^{2}\alpha}\biggr). \end{equation}

\begin{acknowledgements}
The author thanks Professor L. S. Kuz'menkov for fruitful discussions. The work was supported by the Russian
Foundation for Basic Research (grant no. 16-32-00886) and the Dynasty foundation.
\end{acknowledgements}

\end{document}